\documentclass[11pt]{article}

\usepackage[final]{acl}

\usepackage{times}
\usepackage{latexsym}

\usepackage[T1]{fontenc}

\usepackage[utf8]{inputenc}

\usepackage{microtype}

\usepackage{inconsolata}

\usepackage{graphicx}

\usepackage{amsmath}
\usepackage{amssymb}
\usepackage{algorithm}
\usepackage{algpseudocode}
\usepackage{booktabs}
\usepackage{tcolorbox}
\usepackage{float}
\usepackage{multirow}
\usepackage{tabularx}
\usepackage{adjustbox}
\usepackage{pifont}
\usepackage{placeins}

\usepackage[table]{xcolor}
\definecolor{metricgray}{gray}{0.96}

\newcommand{\method}{\textsc{Re-Present}}

\title{\method: Evidence Triangulation across Alternative Candidate-Space Views for Efficient LLM Ensemble Inference}

\author{
 \textbf{Juming Xiong\textsuperscript{1}},
 \textbf{Weixin Liu\textsuperscript{1}},
 \textbf{Kevin Guo\textsuperscript{1}},
 \textbf{Congning Ni\textsuperscript{2}},
 \textbf{Chongyu Qu\textsuperscript{1}},
 \textbf{Junchao Zhu\textsuperscript{1}},
\\
 \textbf{Katherine Brown\textsuperscript{2}},
 \textbf{Chao Yan\textsuperscript{2}},
 \textbf{Bradley Malin\textsuperscript{1,2}},
 \textbf{Zhijun Yin\textsuperscript{1,2}},
\\
\\
 \textsuperscript{1}Vanderbilt University,
 \textsuperscript{2}Vanderbilt University Medical Center
}

\begin{document}
\maketitle

\begin{abstract}
Heterogeneous large language models (LLMs) exhibit complementary strengths, motivating routing and multi-model inference methods that seek to identify reliable predictions for each input. However, exploiting this complementarity remains challenging even when a correct prediction is available within the ensemble. We investigate this challenge across heterogeneous LLM ensembles and find that a simple single-round aggregation rule is highly competitive with learned routing methods, yet substantial oracle headroom remains. Further analysis shows that the evaluated single-round response signals provide limited discriminative power for identifying reliable predictions. In contrast, controlled re-presentations of the same question elicit complementary evidence about prediction reliability. Motivated by these findings, we introduce \method{}, a training-free framework that selectively re-presents disputed questions and directly aggregates candidate-level evidence across alternative views. Across seven multiple-choice benchmarks and two heterogeneous model ensembles, \method{} improves over strong single-round aggregation while offering a favorable accuracy--cost trade-off compared with self-consistency and multi-model debate.
\end{abstract}

\section{Introduction}

Large language models (LLMs) exhibit complementary strengths across tasks and individual examples, motivating routing, ranking, cascading, and multi-model aggregation methods that seek to exploit heterogeneous model capabilities \cite{chen2024frugalgpt,feng2025graphrouter,jiang2023llmblender,zhang2026agentrouter}. A central challenge shared by these approaches is determining which model or prediction should be trusted for a given input. Even when a correct answer is available within an ensemble, identifying and recovering it remains difficult.

Existing routing methods primarily focus on estimating model utility from the input or from available response signals, often relying on supervised training with model-performance data\cite{ong2025routellm,zhang2026agentrouter}. However, model complementarity does not necessarily imply that per-example reliability can be accurately inferred from these observations. Incorrect predictions may receive high confidence, and models with similar aggregate performance may behave differently on individual examples. Consequently, increasingly expressive routing mechanisms may offer limited benefits when the available signals do not sufficiently distinguish reliable predictions from unreliable ones.

We investigate this challenge in heterogeneous LLM ensembles on multiple-choice reasoning tasks. We first find that a simple parameter-free log-linear pool achieves the highest average accuracy among the evaluated single-round routing and ranking methods, yet remains 12.6 and 8.7 percentage points below an ``oracle'' that succeeds whenever any ensemble member answers correctly. Despite this substantial headroom, a learned reliability selector using ensemble-wide single-round response signals provides little improvement over direct aggregation. These findings motivate a complementary perspective: rather than solely developing more sophisticated selection or aggregation algorithms based on fixed single-round observations, heterogeneous ensemble inference may benefit from acquiring additional evidence that better distinguishes competing predictions.

To investigate this possibility, we examine model behavior under controlled re-presentations of the same underlying question. We find that cross-presentation behavior provides reliability information beyond the evaluated single-round response signals. While additional inference through repeated sampling or multi-model interaction can improve prediction accuracy, interactive approaches often incur substantial generation and context overhead\cite{wang2023selfconsistency,du2024debate}. Motivated by these observations, we introduce \method, a training-free framework that selectively re-presents disputed questions under two alternative candidate-space views and directly aggregates the newly elicited candidate-level evidence. Rather than explicitly estimating which model is most reliable or exchanging reasoning trajectories among models, \method{} incorporates evidence across independent presentations into the final prediction. Across seven multiple-choice benchmarks and two heterogeneous model ensembles, the proposed approach improves over strong single-round aggregation and achieves a favorable accuracy--cost trade-off compared with self-consistency and multi-model debate. In summary, our contributions are threefold:
\begin{itemize}
    \item We characterize the gap between practical single-round inference and oracle model complementarity, showing that substantial headroom remains even after strong aggregation.
    \item We investigate whether single-round response signals support reliable sample-level model selection and find that the evaluated signals provide limited discriminative power despite substantial oracle selection headroom.
    \item We show that sample re-presentation elicits complementary reliability evidence and introduce \method{}, a selective, training-free framework that directly integrates candidate-level evidence across alternative views.
\end{itemize}

\section{Related Work}

\paragraph{LLM routing, selection, and aggregation.}
LLM routing exploits heterogeneous model capabilities by selecting models or combining their predictions according to input characteristics and expected performance. Cost-aware cascades sequentially escalate queries across models \cite{chen2024frugalgpt}, while learned routers estimate model utility from preference data or structured representations of queries and candidate models \cite{ong2025routellm,feng2025graphrouter}. AgentRouter further employs a heterogeneous graph neural network trained with performance-derived supervision to produce routing distributions over agents \cite{zhang2026agentrouter}. Adaptive routing methods also jointly consider model selection and test-time sampling budgets \cite{ding2025bestroute}. For multi-turn tasks, MTRouter uses joint history--model embeddings to estimate model utility at each turn \cite{zhang2026mtrouter}. Post-generation approaches instead rank or combine candidate responses, including pairwise ranking in LLM-Blender \cite{jiang2023llmblender} and probability-based aggregation. While our single-round log-linear pooling rule falls into this latter category, we investigate whether additional evidence can help distinguish reliable predictions within heterogeneous ensembles.

\paragraph{Additional test-time inference and interaction.}
Additional inference can improve reasoning through repeated sampling or interaction. Self-consistency aggregates independently sampled reasoning trajectories \cite{wang2023selfconsistency}, while Adaptive-Consistency dynamically allocates samples according to answer agreement \cite{aggarwal2023adaptive}. Soft Self-Consistency further exploits likelihood-based scores rather than relying solely on discrete voting \cite{wang2024softsc}. Confidence-aware selective sampling uses signals from an initial reasoning trajectory to decide whether to trigger additional sampling \cite{xiong2026selectivesampling}. Alternative approaches generate multiple paraphrases of the same question and aggregate their predictions to improve reasoning efficiency \cite{chen2024selfpara}. Multi-agent debate instead allows models to revise predictions after observing other agents' arguments \cite{du2024debate}, with recent methods attempting to mitigate error propagation by filtering unreliable intermediate information \cite{tian2026multiagentdebatememorymasking}. However, multi-model interaction can incur substantial generation and context overhead. Separately, studies of multi-turn clinical question answering show that sequential conversational presentations can degrade accuracy and abstention reliability \cite{guo2026stop}. In contrast, \textsc{Re-Present} acquires additional evidence by changing the candidate spaces only for disputed questions, reducing computation on high-agreement examples.

\paragraph{Reliability under alternative presentations.}
Semantically equivalent rephrasings have been applied to
examine prediction instability on individual examples \cite{yang2024rephrase,faghih2026samequestion}. Multiple-choice LLMs are sensitive to option order and option-label bias \cite{zheng2024robust,pezeshkpour2024options}. Related work in listwise ranking uses permutation self-consistency to mitigate positional bias \cite{tang2024permutation}. 
Building on this observation, we investigate whether controlled changes in the candidate space provide complementary reliability evidence in heterogeneous LLM ensembles. Unlike methods primarily designed to assess presentation robustness or mitigate option-position bias, \textsc{Re-Present} selectively acquires candidate-level evidence from restricted and reordered answer spaces and directly incorporates it into ensemble prediction.

\begin{figure*}[t]
    \centering
    \includegraphics[width=1.0\linewidth] {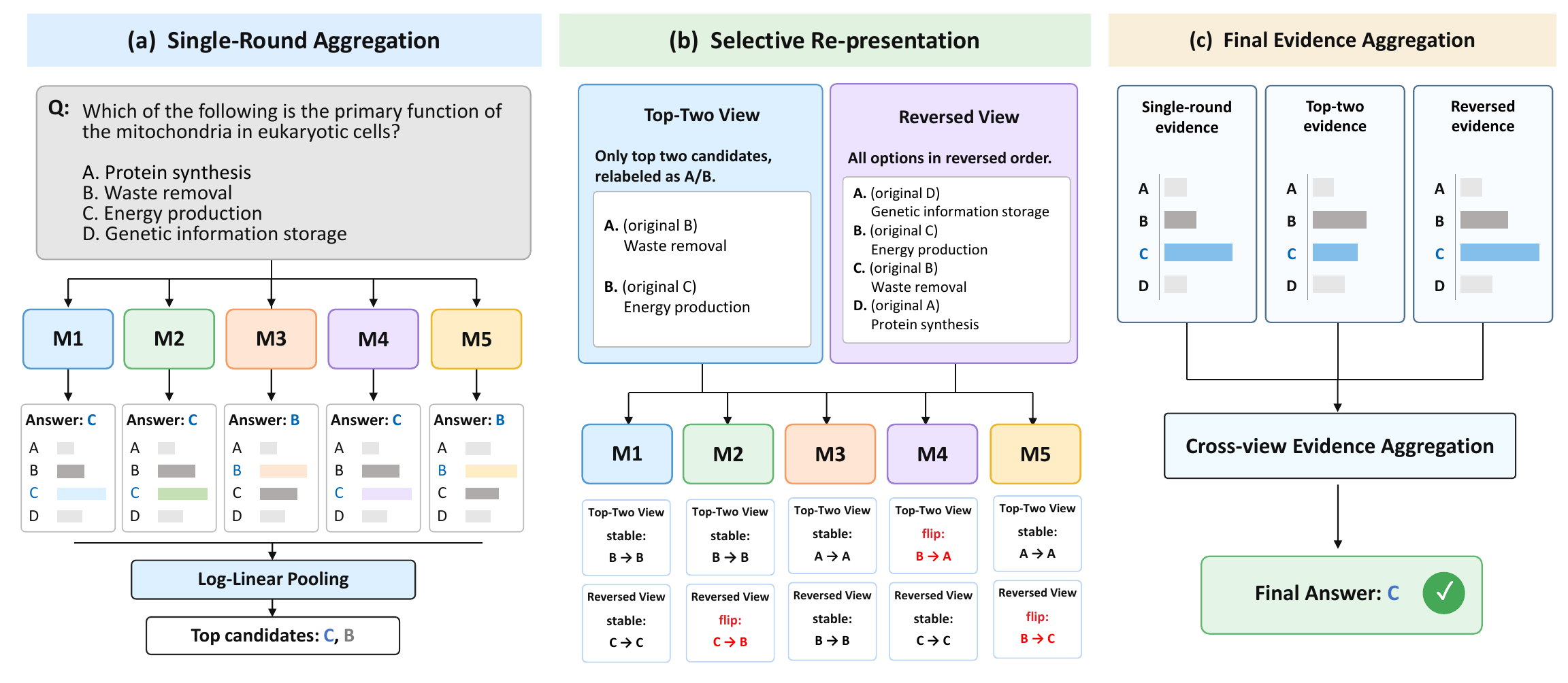}
    \caption{
    \textbf{Overview of RePresent.} The framework first aggregates single-round model evidence, selectively re-presents disputed candidates under alternative views, and combines cross-view evidence for the final prediction.}
    \label{fig:early_exit_framework}
\end{figure*}

\section{Understanding the Routing Gap}
\label{sec:routing_gap}


\subsection{How Effective Is Single-Round Routing?}
\label{sec:single_round_gap}

Table~\ref{tab:single_round} compares standard aggregation rules with learned routing and ranking methods across seven multiple-choice benchmarks. We adopt standard log-linear pooling \citep{Heskes1997Selecting}, which aggregates candidate-level probabilities by summing their log values (see Equation.~\ref{eq:stage1}). This parameter-free rule achieves the highest average accuracy among the evaluated single-round methods.It achieves 80.31\% and 86.67\% accuracy for the 7--9B and 14--32B Ensembles, respectively, leading on all seven benchmarks for the former and five of seven for the latter. However, the ensemble oracle, a hypothetical performance ceiling, which counts a question as correct whenever at least one member answers correctly, achieves 92.93\% and 95.40\% average accuracy for the two ensembles. However, the ensemble \textbf{oracle}, which is correct whenever at least one member answers correctly, achieves hypothetical upper-bound accuracies of 92.93\% and 95.40\% for the two ensembles, respectively. 
The gap is present across all seven benchmarks, indicating that \textbf{the ensemble frequently contains a correct answer even when practical single-round aggregation fails to recover it \textsc{(Key Finding 1).}}

\begin{table*}[t]
\centering
\small
\setlength{\tabcolsep}{4.0pt}
\begin{tabular}{lcccccccc}
\toprule
\textbf{Method} & \textbf{MedQA} & \textbf{MedMCQA} & \textbf{MMLU} & \textbf{ARC-C} & \textbf{OpenBookQA} & \textbf{MathQA} & \textbf{HeadQA} & \textbf{Avg.} \\
\midrule
\multicolumn{9}{l}{\textit{7--9B Ensemble}} \\
\midrule
Average & 60.9 & 57.6 & 78.1 & 88.3 & 85.0 & 71.7 & 68.3 & 72.83 \\
Best Single LLM & 66.8 & 61.5 & 78.8 & 91.0 & 86.0 & 85.0 & 75.1 & 77.74 \\
Majority Vote & 66.7 & 62.5 & 81.5 & 92.3 & 90.4 & 83.5 & 74.4 & 78.76 \\
LLM-Blender & 57.6 & 57.3 & 78.1 & 89.1 & 84.9 & 74.3 & 67.1 & 72.63 \\
GraphRouter & 59.0 & 57.2 & 79.4 & 90.0 & 88.6 & 66.5 & 66.1 & 72.40 \\
AgentRouter & 66.9 & 62.0 & 81.8 & 92.0 & 90.3 & 84.6 & 75.1 & 78.97 \\
Log-Linear Pooling & \textbf{69.9} & \textbf{64.3} & \textbf{82.3} & \textbf{93.0} & \textbf{91.3} & \textbf{85.6} & \textbf{75.8} & \textbf{80.31} \\
\midrule
Oracle & 87.5 & 86.0 & 94.4 & 98.1 & 98.0 & 95.3 & 91.2 & 92.93 \\
\midrule
\multicolumn{9}{l}{\textit{14--32B Ensemble}} \\
\midrule
Average & 73.9 & 65.6 & 85.5 & 91.9 & 89.7 & 82.6 & 78.3 & 81.06 \\
Best Single LLM & \textbf{81.6} & 69.0 & \textbf{90.3} & 94.7 & 93.8 & 86.0 & 81.8 & 85.31 \\
Majority Vote & 80.4 & 71.6 & 88.9 & 96.4 & 95.2 & 88.4 & 83.6 & 86.36 \\
LLM-Blender & 72.8 & 67.9 & 85.2 & 94.1 & 91.9 & 81.1 & 78.7 & 81.67 \\
GraphRouter & \textbf{81.6} & 69.3 & 89.8 & 94.3 & 93.5 & 84.9 & 81.9 & 85.04 \\
AgentRouter & 79.9 & 71.9 & 88.6 & 96.3 & 94.5 & 88.3 & 82.4 & 85.98 \\
Log-Linear Pooling & 79.7 & \textbf{72.2} & 88.7 & \textbf{96.5} & \textbf{95.5} & \textbf{90.1} & \textbf{84.0} & \textbf{86.67} \\
\midrule
Oracle & 94.2 & 89.4 & 97.4 & 98.3 & 98.7 & 95.5 & 94.3 & 95.40 \\
\bottomrule
\end{tabular}
\caption{Per-dataset test accuracy (\%) of single-round routing and aggregation methods. Each dataset contains 1,000 test questions, and Avg.\ denotes the unweighted mean across seven datasets. Average denotes the mean accuracy of the five individual models, while Best Single LLM denotes the highest-performing individual model for each dataset. AgentRouter results are averaged over three seeds. The oracle counts a question as correct whenever at least one ensemble member produces the correct answer. Bold indicates the highest non-oracle accuracy in each column, including ties.}
\label{tab:single_round}
\end{table*}


\subsection{Can Single-Round Response Signals Identify Reliable Models?}
\label{sec:single_round_selection}

Motivated by this \textbf{oracle gap}, we examine whether single-round response signals can reliably identify the best model for each question. We train a logistic regression classifer to estimate prediction correctness for each model–question pair and select the model with the highest estimated probability at test time. We consider two feature sets: model-specific response signals alone and an expanded set incorporating ensemble-level agreement and aggregation statistics. 
Table~\ref{tab:single_round_selection} reports accuracy on disputed questions. For the 7--9B Ensemble, the classifier using all single-round signals achieves 54.7\% accuracy, compared with 54.9\% for log-linear pooling. For the 14--32B Ensemble, the corresponding accuracies are 52.6\% and 53.1\%. Using only model-specific signals yields 49.8\% and 53.1\%, respectively. Thus, the evaluated learned selectors do not consistently outperform direct aggregation, despite oracle selection accuracies of 89.6\% and 91.8\%.

\begin{table}[t]
\centering
\small
\setlength{\tabcolsep}{5pt}
\begin{tabular}{lcc}
\toprule
\textbf{Decision Rule} & \textbf{7--9B} & \textbf{14--32B} \\
\midrule
Random LLM & 40.2 & 40.5 \\
Best Fixed Model & 49.1 & 45.9 \\
Highest-Confidence LLM & 48.5 & 48.2 \\
LR (Per-Model Signals) & 49.8 & \textbf{53.1} \\
LR (All Single-Round Signals) & \underline{54.7} & 52.6 \\
Majority Vote & 49.4 & 51.1 \\
Log-Linear Pooling & \textbf{54.9} & \textbf{53.1} \\
\midrule
Oracle Selection & 89.6 & 91.8 \\
\bottomrule
\end{tabular}
\caption{Accuracy (\%) on disputed questions using single-round evidence. LR uses model-specific or ensemble-level features. Best Fixed Model selects the top-performing model on the training data. Oracle Selection assumes a correct answer whenever any model is correct. Best non-oracle results are bolded; second-best results are underlined.}
\label{tab:single_round_selection}
\end{table}

The classifiers tend to favor globally strong models rather than reliably identifying the best model for each question. For example, Llama-8B and GPT-OSS-20B are selected for 52\% and 59\% of disputed questions in their respective ensembles. Even with ensemble-level information, the evaluated single-round features achieve only modest within-question AUROCs of 0.67 and 0.63. These findings suggest \textbf{limited discriminative power for question-specific model selection, motivating the acquisition of additional reliability evidence \textsc{(Key Finding 2)}.}



\subsection{Does Re-presentation Provide Additional Reliability Evidence?}
\label{sec:representation_signal}

To better distinguish competing predictions, we examine two re-presentations of disputed questions: a \textbf{top-two view} retaining the two highest-scoring candidates in their original relative order, and a \textbf{reversed view} reversing the order of all answer options. Each model answers both views independently, without access to prior predictions or other models' responses. We then examine whether prediction stability, defined as selecting the same candidate after re-presentation, is associated with correctness. For the 7--9B Ensemble, stable predictions are more accurate than flipped predictions under both the top-two (51.6\% vs. 25.7\%) and reversed (53.6\% vs. 27.6\%) views. Similar gaps of 24.1 and 24.3 percentage points are observed for the 14--32B Ensemble Appendix~\ref{app:stability}.

Re-presentation provides complementary evidence for predicting correctness, improving AUROC from 0.722 to 0.745 for the 7--9B Ensemble and from 0.700 to 0.724 for the 14--32B Ensemble when combined with single-round features. We further examine high-confidence predictions, defined as those with single-round confidence at or above the median among model--question predictions on disputed questions (0.982 and 0.998 for the 7--9B and 14--32B Ensembles, respectively).Among these predictions, those stable across both views are substantially more accurate than those that change, with accuracies of 64.3\% versus 36.4\% for the 7--9B Ensemble and 61.5\% versus 34.1\% for the 14--32B Ensemble. These findings demonstrate that \textbf{re-presentation provides additional reliability evidence beyond single-round response signals \textsc{(Key Finding 3)}.}



\begin{table}[t]
\centering
\small
\begin{tabular}{lcc}
\toprule
\textbf{Reliability Signal} & \textbf{7--9B} & \textbf{14--32B} \\
\midrule
Single-Round Response Signals & 0.722 & 0.700 \\
Re-presentation Signals & 0.708 & 0.692 \\
Combined & \textbf{0.745} & \textbf{0.724} \\
\bottomrule
\end{tabular}
\caption{AUROC for predicting individual model correctness on disputed questions. Combining single-round features with cross-presentation behavior improves discrimination in both ensembles.}
\label{tab:combined_reliability}
\end{table}





\section{\method: Selective Evidence Triangulation}
\label{sec:method}


Motivated by the three key findings, we introduce \method, a training-free framework that selectively acquires additional evidence to resolve disagreements in heterogeneous LLM ensembles. In Stage~1, models independently answer the original question, and their candidate-level distributions are combined using log-linear pooling. Questions with strong agreement terminate at this stage. For disputed questions, Stage~2 independently evaluates two alternative candidate-space views and aggregates their evidence with the original predictions. The framework requires neither a learned router nor communication among models (Figure~\ref{fig:early_exit_framework}).

\subsection{Stage 1: Single-Round Log-Linear Pooling}

Let $\mathcal{O}$ denote the set of answer candidates and $\{1,\ldots,K\}$ index the ensemble members. Each model independently generates a response and commits to an answer. We extract the logits corresponding to the candidate options at the final-answer decision point and normalize them using softmax to obtain a candidate-level probability distribution $p_i^0(o)$ for each model $i$. When a valid final answer cannot be parsed, we assign a uniform distribution over $\mathcal{O}$.

We aggregate these distributions using log-linear pooling:
\begin{equation}
s_1(o)
=
\sum_{i=1}^{K}
\log p_i^0(o).
\label{eq:stage1}
\end{equation}
The Stage~1 prediction is
\begin{equation}
\hat{o}_1
=
\arg\max_{o\in\mathcal{O}} s_1(o).
\end{equation}
This parameter-free aggregation serves as the single-round backbone of \method{}.

\subsection{Selective Re-presentation}

We allocate additional inference according to disagreement among the models' committed single-round answers. Let $a_i\in\mathcal{O}$ denote the answer selected by model $i$. We define the size of the largest agreement group as
\begin{equation}
m
=
\max_{o\in\mathcal{O}}
\sum_{i=1}^{K}
\mathbb{I}[a_i=o].
\end{equation}
Stage~2 is invoked when
\begin{equation}
m \le K-2.
\label{eq:trigger}
\end{equation}
For $K=5$, this criterion skips unanimous and $4$--$1$ configurations while allocating additional inference to questions with greater disagreement.

For each selected question, we construct two alternative views. The \emph{top-two} view retains the two candidates with the highest Stage~1 scores, preserves their relative order in the original question, and relabels them as A/B. The \emph{reversed} view retains all candidates but reverses their presentation order and assigns new option labels accordingly. Each model answers the two views independently, without observing other models' responses or its own previous prediction.

\subsection{Final Cross-View Evidence Aggregation}

Let $\mathcal{T}\subseteq\mathcal{O}$ contain the two highest-scoring candidates under Stage~1. For each re-presentation view $v\in\mathcal{V}=\{\mathrm{top2},\mathrm{rev}\}$, we obtain a candidate-level probability distribution $p_i^v(o)$ from each model using the same logit-based procedure as in Stage~1. We map these distributions back to their original candidate identities and normalize them over $\mathcal{T}$, yielding $\tilde{p}_i^v(o)$.

We aggregate the newly elicited candidate-level evidence as
\begin{equation}
s_2(o)
=
\sum_{v\in\mathcal{V}}
\sum_{i=1}^{K}
\log \tilde{p}_i^v(o),
\qquad o\in\mathcal{T}.
\label{eq:stage2}
\end{equation}
For disputed questions, the final prediction is
\begin{equation}
\hat{o}
=
\arg\max_{o\in\mathcal{T}}
\left[s_1(o)+w\,s_2(o)\right],
\label{eq:represent}
\end{equation}
while questions not selected for Stage~2 retain $\hat{o}_1$.

The scalar $w$ controls the contribution of the additional evidence and is selected on a disjoint training split from $\{0.1,0.25,0.5,1,2,4\}$. Both ensembles select $w=1$. Rather than converting cross-presentation behavior into an explicit model-reliability estimate, \method{} directly aggregates the newly elicited candidate-level distributions. This preserves information about which answers are supported across presentations without requiring a learned router or reliability predictor.

\section{Results}
\label{sec:results}

\begin{figure*}[t]
    \centering
    \includegraphics[width=1.0\linewidth]{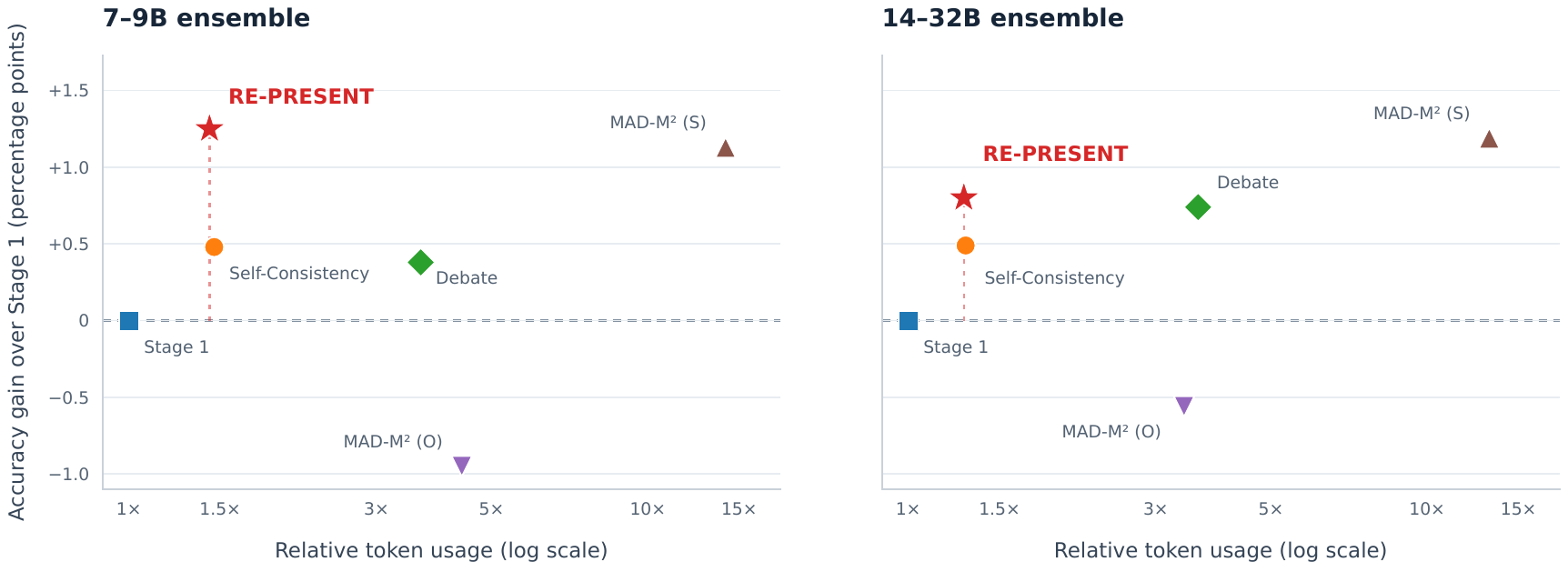}
    \caption{
        Accuracy--token trade-off across two heterogeneous model ensembles.
        Each point represents average accuracy and total token usage,
        normalized by the initial five-model inference. \method{} achieves competitive accuracy with substantially lower token usage than the interactive baselines.
    }
    \label{fig:accuracy_cost}
\end{figure*}

We evaluate \method{} against additional-inference strategies that improve predictions through repeated sampling or multi-model interaction. Self-consistency aggregates additional independently generated responses, while Debate and MAD-M$^2$ allow models to revise predictions through interaction. Figure~\ref{fig:accuracy_cost} reports per-dataset accuracy and additional prompt and generated tokens relative to the initial five-model single-round inference. Self-consistency and \method{} use the same disagreement-based selective trigger, whereas the main Debate and MAD-M$^2$ configurations apply additional inference to all questions.

\paragraph{Overall accuracy.}
\method{} improves over single-round log-linear pooling in both ensembles, increasing average accuracy from 80.31\% to 81.51\% for the 7--9B Ensemble and from 86.67\% to 87.46\% for the 14--32B Ensemble. These correspond to gains of 1.14 and 0.76 percentage points, respectively. Among the evaluated additional-inference methods, \method{} achieves the highest average accuracy for the 7--9B Ensemble, slightly exceeding MAD-M$^2$(S) at 81.44\%. For the 14--32B Ensemble, MAD-M$^2$(S) achieves the highest average accuracy of 87.86\%, while \method{} reaches 87.46\%, a difference of 0.40 percentage points.

\paragraph{Accuracy--cost trade-off.}
The accuracy improvements of \method{} are obtained with relatively low additional token overhead. For the 7--9B Ensemble, \method{} requires 60\% additional prompt tokens and 32\% additional generated tokens, compared with 63\% and 34\% for selective self-consistency. Despite similar token budgets, \method{} achieves higher average accuracy (81.51\% versus 80.79\%). The same pattern holds for the 14--32B Ensemble, where \method{} achieves 87.46\% accuracy with 35\% additional prompt tokens and 25\% additional generated tokens, compared with 87.16\% accuracy and 37\% and 26\% additional tokens for self-consistency.

Interactive baselines incur substantially greater additional token overhead under their full-inference configurations. For example, MAD-M$^2$(S) requires 2142\% additional prompt tokens and 764\% additional generated tokens for the 7--9B Ensemble, compared with 60\% and 32\% for \method{}. For the 14--32B Ensemble, the corresponding costs are 3098\% and 460\%, compared with 35\% and 25\%. Standard Debate also incurs substantially higher prompt-token overhead, while achieving lower average accuracy than \method{} in both ensembles.

\section{Analysis}
\label{sec:analysis}


\subsection{Why Use Two Re-presentations?}

The two re-presentation views provide complementary evidence for resolving disputed predictions. As shown in Table~\ref{tab:view_ablation}, incorporating only the reversed view improves average accuracy from 80.31\% to 81.19\% for the 7--9B Ensemble and from 86.67\% to 87.09\% for the 14--32B Ensemble. Using only the top-two view yields 80.91\% and 87.16\%, respectively. Combining both views achieves the highest accuracy in both ensembles, reaching 81.51\% and 87.46\%.

\begin{table}[t]
\centering
\small
\begin{tabular}{lcc}
\toprule
\textbf{Evidence} & \textbf{7--9B} & \textbf{14--32B} \\
\midrule
Stage~1 Only & 80.31 & 86.67 \\
+ Reversed View & 81.19 & 87.09 \\
+ Top-Two View & 80.91 & 87.16 \\
+ Both Views & \textbf{81.51} & \textbf{87.46} \\
\bottomrule
\end{tabular}
\caption{Average accuracy (\%) when incorporating different re-presentation views into Stage~1 evidence.}
\label{tab:view_ablation}
\end{table}

The benefit of combining both views is consistent with their partially distinct effects on model predictions. The top-two view restricts the candidate space, whereas the reversed view preserves all candidates but changes their presentation order. Their error patterns are only moderately correlated, with coefficients of $\phi=0.34$ and $\phi=0.39$ for the 7--9B and 14--32B Ensembles, respectively (Appendix~\ref{app:controls}). Thus, the two views provide complementary observations, although neither transformation is individually necessary for eliciting additional reliability evidence.

\subsection{Where Does Re-presentation Help?}

The benefit of \method{} varies with the degree of single-round disagreement. For the 7--9B Ensemble, accuracy on $3$--$2$ voting splits increases from 57.1\% to 63.5\%, while accuracy on $2$--$2$--$1$ splits increases from 45.6\% to 50.5\%. The corresponding improvements for the 14--32B Ensemble are from 57.5\% to 63.2\% and from 43.6\% to 47.3\%.

By contrast, high-agreement configurations exhibit little benefit from additional inference. As detailed in Appendix~\ref{app:trigger_analysis}, re-presentation produces negligible changes under unanimity and little or no net improvement on $4$--$1$ splits. These observations support the disagreement-based trigger in Eq.~\ref{eq:trigger}, which allocates additional inference to questions where competing predictions remain unresolved while avoiding unnecessary model calls on high-agreement examples.

\subsection{Why Not Explicitly Learn Whom to Trust?}

Finding~2 shows that reliability estimation using the evaluated single-round signals does not consistently outperform direct aggregation. We next examine whether explicitly estimating model reliability becomes more effective after acquiring re-presentation evidence.

We train logistic-regression reliability classifiers using both single-round and re-presentation signals and evaluate two decision strategies on disputed questions. The first selects the model with the highest estimated probability of correctness, while the second weights model votes by their estimated reliability. We compare these approaches with directly aggregating the candidate-level distributions elicited by \method{}.

\begin{table}[t]
\centering
\small
\begin{tabular}{lcc}
\toprule
\textbf{Decision Rule} & \textbf{7--9B} & \textbf{14--32B} \\
\midrule
Stage~1 Log-Linear Pooling & 54.9 & 53.1 \\
Select Predicted-Best Model & 56.5 & 53.2 \\
Reliability-Weighted Vote & 54.9 & 51.7 \\
\method{} Pooling & \textbf{58.9} & \textbf{57.5} \\
\bottomrule
\end{tabular}
\caption{Accuracy (\%) on disputed questions using different strategies for incorporating re-presentation evidence. Learned reliability strategies use both single-round and re-presentation signals, while \method{} directly aggregates candidate-level distributions.}
\label{tab:reliability_vs_pooling}
\end{table}

As shown in Table~\ref{tab:reliability_vs_pooling}, explicit model selection improves over Stage~1 pooling for the 7--9B Ensemble, reaching 56.5\%, but provides little benefit for the 14--32B Ensemble. Reliability-weighted voting similarly offers no consistent advantage. In contrast, directly pooling candidate-level evidence achieves 58.9\% and 57.5\% accuracy across the two ensembles.

These results suggest that re-presentation evidence is more effectively used at the candidate level than through the evaluated scalar model-reliability estimates. In particular, reducing cross-presentation observations to a single reliability score per model may discard information about which candidates receive support across alternative views. Direct pooling avoids this intermediate reduction and provides a simple, training-free way to incorporate the additional evidence.

\subsection{Is the Gain Explained by More Computation?}

A natural question is whether the improvements arise from re-presentation itself or simply from allocating additional inference. To examine this distinction, we compare alternative presentations against repeated inference under the original presentation, using similar additional token budgets.

As shown in Appendix~\ref{app:controls}, an additional greedy inference call under the same presentation improves average accuracy by 0.37 and 0.26 percentage points for the 7--9B and 14--32B Ensembles, respectively. Additional temperature-based sampling also provides smaller or inconsistent gains. In contrast, combining the top-two and reversed views improves accuracy by 1.14 and 0.76 percentage points, respectively.

Additional controls with random option permutations and cyclic shifts further show that the benefit is not unique to reversing the answer order. Together, these comparisons indicate that the improvements cannot be attributed solely to additional token generation: changing the candidate presentation provides useful evidence beyond that obtained from repeated inference under an unchanged presentation.

\section{Experimental Setup}
\label{sec:setup}

\paragraph{Benchmarks.}
We evaluate on seven multiple-choice question-answering benchmarks: MedQA \cite{jin2021medqa}, MedMCQA \cite{pal2022medmcqa}, MMLU \cite{hendrycks2021mmlu}, ARC-Challenge (ARC-C) \cite{clark2018arc}, OpenBookQA \cite{mihaylov2018openbookqa}, MathQA \cite{amini2019mathqa}, and HeadQA \cite{vilares2019headqa}. Each benchmark contains 1,000 test questions and a disjoint set of 1,000 questions used for training reliability classifiers, selecting hyperparameters, and configuring learned baselines. We report accuracy on each benchmark and the unweighted average across all seven datasets.

\paragraph{Model ensembles.}
We construct two non-overlapping heterogeneous ensembles, each consisting of five LLMs. The 7--9B Ensemble comprises Gemma2-9B \cite{gemma2024gemma2}, Llama-8B \cite{grattafiori2024llama3}, Ministral-8B \cite{mistral2025ministral3}, Qwen2.5-7B \cite{yang2025qwen25}, and Qwen3-8B \cite{yang2025qwen3}. The 14--32B Ensemble comprises GPT-OSS-20B \cite{openai2025gptoss}, Gemma3-27B \cite{gemma2025gemma3}, Ministral-14B \cite{mistral2025ministral3}, Qwen3-14B, and Qwen3-32B \cite{yang2025qwen3}. These ensembles represent different model families and parameter scales, allowing us to examine whether the findings generalize across distinct model pools.

\paragraph{Inference protocol.}
Each model independently generates a short chain-of-thought response using greedy decoding. We extract candidate-level probability distributions and aggregate them using the log-linear pooling rule in Eq.~\ref{eq:stage1}. For \method{}, additional inference is performed only on questions satisfying the disagreement criterion in Eq.~\ref{eq:trigger}. This identifies 2,027 disputed test questions for the 7--9B Ensemble and 1,196 for the 14--32B Ensemble across the seven benchmarks. Each selected question is independently evaluated under the top-two and reversed views, without inter-agent communication or access to previous predictions.

\paragraph{Baselines.}
For single-round inference, we compare against the mean accuracy of individual ensemble members (Average), the best individual model for each dataset (Best Single LLM), Majority Vote, LLM-Blender \cite{jiang2023llmblender}, GraphRouter \cite{feng2025graphrouter}, AgentRouter \cite{zhang2026agentrouter}, and Log-Linear Pooling. AgentRouter results are averaged over three random seeds. For additional inference, we evaluate self-consistency \cite{wang2023selfconsistency}, multi-agent Debate \cite{du2024debate}, and MAD-M$^2$ \cite{tian2026multiagentdebatememorymasking} with subjective (S) and objective (O) masking. Self-consistency and \method{} use the same disagreement-based selective trigger, whereas the main Debate and MAD-M$^2$ configurations perform additional inference on all questions.



\section{Discussion and Conclusion}

This work investigates the gap between model complementarity and the ability to exploit it in heterogeneous LLM ensembles. Across seven multiple-choice benchmarks and two model ensembles, we find that strong single-round aggregation remains substantially below oracle performance, while the evaluated learned selectors provide limited improvements from single-round response signals. These observations motivate a complementary perspective on heterogeneous ensemble inference: improving the evidence available for prediction selection, rather than solely refining the decision rule.

We show that controlled re-presentation elicits complementary reliability evidence beyond the evaluated single-round signals. Building on this finding, we introduce \method{}, which selectively re-presents disputed questions and directly aggregates the resulting candidate-level evidence. Without requiring a learned router or inter-agent communication, \method{} improves over strong single-round aggregation and achieves a favorable accuracy--cost trade-off compared with the evaluated additional-inference methods.

More broadly, our findings suggest that acquiring informative observations offers a complementary direction for heterogeneous LLM inference, alongside improving model selection and aggregation strategies.

\section*{Limitations}

Our study focuses on multiple-choice question answering, where candidate-level probabilities and alternative presentations are readily available. Extending re-presentation to open-ended tasks remains a direction for future research. We evaluate two heterogeneous five-model ensembles across seven benchmarks, with future work exploring additional model configurations and task settings. Although re-presentation incurs additional inference costs, it offers a favorable accuracy–cost trade-off compared with more expensive multi-round methods. Finally, prediction stability is associated with correctness but does not guarantee it. Future work could further investigate the underlying mechanisms and develop adaptive strategies for selecting views and allocating inference budgets.

\section*{Ethical Considerations}

This study evaluates LLM ensembles using publicly available multiple-choice benchmarks. Although
re-presentation improves prediction accuracy, stable predictions do not guarantee correctness
and should not guide high-stakes decisions without expert oversight, particularly in medical settings. Additional inference also incurs computational and environmental costs, which are mitigated through selective execution.

\bibliography{custom}

\newpage

\appendix

\section{Detailed Cross-Presentation Stability Analysis}
\label{app:stability}

We provide additional analyses supporting the relationship between cross-presentation stability and single-round correctness discussed in Section~\ref{sec:representation_signal}. All analyses are restricted to disputed questions, with each model--question pair treated as one prediction instance. A prediction is considered \emph{stable} if the model selects the same original answer candidate after re-presentation, and a \emph{flip} otherwise.

\paragraph{Overall stability.}
Table~\ref{tab:stability_main} reports single-round correctness conditioned on prediction stability under the two re-presentation views. Across both ensembles, stable predictions are substantially more likely to be correct than flipped predictions. The stable--flip accuracy gaps range from approximately 24 to 26 percentage points, suggesting that cross-presentation stability provides a useful indicator of prediction reliability.


\begin{table}[!htbp]
\centering
\small
\setlength{\tabcolsep}{2.5pt}
\resizebox{\columnwidth}{!}{%
\begin{tabular}{@{}llccccc@{}}
\toprule
& & \multicolumn{2}{c}{\textbf{Stable}} &
\multicolumn{2}{c}{\textbf{Flip}} & \\
\cmidrule(lr){3-4}
\cmidrule(lr){5-6}
\textbf{Ens.} & \textbf{View} &
\textbf{$n$} & \textbf{Acc.} &
\textbf{$n$} & \textbf{Acc.} &
\textbf{Gap} \\
\midrule
7--9B
& Top-Two
& 5692 & 51.6 & 4443 & 25.7 & +26.0 \\
& Reversed
& 4926 & 53.6 & 5209 & 27.6 & +26.0 \\
\midrule
14--32B
& Top-Two
& 3424 & 50.8 & 2556 & 26.7 & +24.1 \\
& Reversed
& 3067 & 52.4 & 2913 & 28.0 & +24.3 \\
\bottomrule
\end{tabular}%
}
\caption{First-pass correctness (\%) by cross-presentation stability on disputed model--question pairs. $n$ denotes the number of model--question pairs in each group. Gap denotes the stable--flip accuracy difference in percentage points.}
\label{tab:stability_main}
\end{table}

Considering both views jointly provides a finer distinction. For the 7--9B Ensemble, single-round correctness is 60.5\% when predictions remain stable under both views, 37.3\% when stable under exactly one view, and 20.3\% when stable under neither. The corresponding accuracies for the 14--32B Ensemble are 58.6\%, 35.7\%, and 22.7\%, respectively. This ordering indicates that stability across multiple views provides a more discriminative descriptive signal than stability under a single view.

\subsection{Stability Across Datasets}

Table~\ref{tab:stability_dataset} reports the association between reversed-view stability and single-round correctness for each benchmark. Positive stable--flip gaps are observed across all seven datasets in both ensembles, although the magnitude varies substantially.

\begin{table}[t]
\centering
\small
\setlength{\tabcolsep}{6pt}
\begin{tabular}{lrr|rr}
\toprule
&
\multicolumn{2}{c|}{\textbf{7--9B Ensemble}} &
\multicolumn{2}{c}{\textbf{14--32B Ensemble}} \\
\textbf{Dataset}
& \textbf{Gap}
& \textbf{AUROC}
& \textbf{Gap}
& \textbf{AUROC} \\
\midrule
MedQA       & +19.8 & 0.629 & +21.9 & 0.625 \\
MedMCQA     &  +8.7 & 0.558 & +17.0 & 0.609 \\
MMLU        & +29.1 & 0.693 & +28.4 & 0.671 \\
ARC-C       & +41.4 & 0.725 & +42.4 & 0.745 \\
OpenBookQA  & +38.5 & 0.739 & +35.4 & 0.695 \\
MathQA      & +42.4 & 0.733 & +35.1 & 0.692 \\
HeadQA      & +25.5 & 0.656 & +18.8 & 0.626 \\
\bottomrule
\end{tabular}
\caption{Per-dataset association between reversed-view stability and single-round correctness. Gap denotes stable accuracy minus flip accuracy in percentage points. AUROC evaluates the binary stability indicator as a predictor of single-round correctness.}
\label{tab:stability_dataset}
\end{table}

The association is particularly pronounced on ARC-C, OpenBookQA, and MathQA, where stable--flip gaps exceed 35 percentage points for both ensembles. Smaller gaps are observed on MedMCQA, suggesting that the discriminative value of stability varies across tasks. Nevertheless, the direction of the association remains consistent across all evaluated benchmarks.

\subsection{Stability Across Models}

We further examine whether the association between stability and correctness is concentrated in particular model families. Table~\ref{tab:stability_model} reports per-model results under the reversed view.

\begin{table}[t]
\centering
\small
\begin{tabular}{lrr}
\toprule
\textbf{Model} & \textbf{Gap} & \textbf{AUROC} \\
\midrule
\multicolumn{3}{l}{\textit{7--9B Ensemble}} \\
Gemma2-9B       & +24.3 & 0.671 \\
Llama3.1-8B     & +23.7 & 0.658 \\
Ministral3-8B   & +38.7 & 0.711 \\
Qwen2.5-7B      & +14.0 & 0.610 \\
Qwen3-8B        & +26.1 & 0.668 \\
\midrule
\multicolumn{3}{l}{\textit{14--32B Ensemble}} \\
GPT-OSS-20B     & +35.9 & 0.678 \\
Gemma3-27B      & +16.5 & 0.601 \\
Ministral3-14B  & +13.9 & 0.599 \\
Qwen3-14B       & +25.0 & 0.657 \\
Qwen3-32B       & +27.3 & 0.666 \\
\bottomrule
\end{tabular}
\caption{Per-model association between reversed-view stability and single-round correctness on disputed questions. Gap denotes stable accuracy minus flip accuracy in percentage points, while AUROC evaluates stability as a binary correctness indicator.}
\label{tab:stability_model}
\end{table}

All ten models exhibit positive stable--flip gaps, although the magnitude varies across model families. This suggests that the association between cross-presentation stability and correctness is not driven by a single model, while also indicating that stability may have different discriminative strengths across ensemble members.

\subsection{Stability Beyond Single-Round Confidence}

Finally, we examine whether cross-presentation stability remains informative within different single-round confidence groups. Table~\ref{tab:confidence_stability} reports correctness stratified by confidence and joint stability across the top-two and reversed views.

\begin{table}[t]
\centering
\small
\begin{tabular}{llrr}
\toprule
\textbf{Confidence} &
\textbf{Behavior} &
\textbf{7--9B} &
\textbf{14--32B} \\
\midrule
High & Stable & 64.3 & 61.5 \\
High & Flip   & 36.4 & 34.1 \\
Low  & Stable & 53.8 & 52.8 \\
Low  & Flip   & 24.5 & 26.6 \\
\bottomrule
\end{tabular}
\caption{Single-round correctness (\%) stratified by confidence and joint cross-presentation stability. Stable indicates agreement with the original prediction under both views, while Flip indicates a change under at least one view.}
\label{tab:confidence_stability}
\end{table}

Among high-confidence predictions, those stable under both views achieve substantially higher correctness than those that change under at least one view: 64.3\% versus 36.4\% for the 7--9B Ensemble and 61.5\% versus 34.1\% for the 14--32B Ensemble. Similar differences appear in the low-confidence group. These results support the conclusion that cross-presentation behavior provides useful reliability information beyond a coarse distinction based on single-round confidence.


\section{Selective Interactive Baselines}
\label{app:selective_baselines}

The main results in Figure~\ref{fig:accuracy_cost} compare selective self-consistency and \method{} with interactive baselines that perform additional inference on all questions. To separate the effect of inference allocation from the inference mechanism, we additionally evaluate Debate and MAD-M$^2$ using the same disagreement-based selective trigger as \method{} (Eq.~\ref{eq:trigger}). Additional inference is invoked only when the largest agreement group contains at most three of the five ensemble members. All other questions retain the Stage~1 log-linear pooling prediction.

Table~\ref{tab:selective_baselines} reports average accuracy and token usage relative to the initial five-model inference, including Stage~1. Under this normalization, $1.00\times$ corresponds to the token volume of the initial five-model pass.

\begin{table}[t]
\centering
\small
\setlength{\tabcolsep}{3.5pt}
\begin{tabular}{lcccc}
\toprule
\textbf{Method} & \textbf{Avg.} &
\multicolumn{3}{c}{\textbf{Token Usage ($\times$)}} \\
\cmidrule(lr){3-5}
& & \textbf{Input} & \textbf{Output} & \textbf{Total} \\
\midrule
\multicolumn{5}{l}{\textit{7--9B Ensemble}} \\
\midrule
Selective Debate
& 81.06 & 2.77 & \textbf{1.26} & 1.86 \\
Selective MAD-M$^2$(S)
& \textbf{81.53} & 7.44 & 3.92 & 5.32 \\
Selective MAD-M$^2$(O)
& 79.93 & 2.80 & 1.56 & 2.05 \\
Selective Self-Consistency
& 80.79 & \underline{1.63} & 1.34 & \underline{1.46} \\
\method{}
& \underline{81.51} & \textbf{1.60} &
\underline{1.32} & \textbf{1.43} \\
\midrule
\multicolumn{5}{l}{\textit{14--32B Ensemble}} \\
\midrule
Selective Debate
& \underline{87.49} & 2.38 & \textbf{1.18} & 1.52 \\
Selective MAD-M$^2$(S)
& \textbf{87.81} & 6.77 & 2.25 & 3.54 \\
Selective MAD-M$^2$(O)
& 86.47 & 2.19 & \underline{1.22} & 1.50 \\
Selective Self-Consistency
& 87.16 & \underline{1.37} & 1.26 &
\underline{1.29} \\
\method{}
& 87.46 & \textbf{1.35} & 1.25 &
\textbf{1.28} \\
\bottomrule
\end{tabular}
\caption{Average accuracy (\%) and token usage under the same disagreement-based selective trigger. Token usage includes the initial inference and is normalized by the initial five-model pass ($1.00\times$). Input, Output, and Total denote input, generated, and combined token volumes, respectively. Bold and underline indicate the best and second-best values within each ensemble, respectively.}
\label{tab:selective_baselines}
\end{table}

Under the same selective trigger, \method{} achieves competitive accuracy with substantially lower total token usage than the interactive baselines. For the 7--9B Ensemble, \method{} reaches 81.51\% accuracy at $1.43\times$ total tokens, closely matching selective MAD-M$^2$(S) at 81.53\% accuracy but requiring only a fraction of its $5.32\times$ token usage. Selective Debate achieves 81.06\% accuracy with $1.86\times$ total tokens.

For the 14--32B Ensemble, \method{} achieves 87.46\% accuracy at $1.28\times$ total tokens. Selective Debate reaches a similar accuracy of 87.49\% but uses $1.52\times$ total tokens, while selective MAD-M$^2$(S) achieves the highest accuracy of 87.81\% at $3.54\times$ token usage.

These results demonstrate that the favorable accuracy--token trade-off of \method{} is not solely due to selective inference. Under the same triggering policy, independently elicited candidate-level evidence achieves competitive accuracy while using fewer total tokens than the evaluated interactive inference methods.


\section{Additional Re-presentation Controls}
\label{app:controls}

We conduct controlled experiments to examine whether the improvements of \method{} arise primarily from additional inference or from changes to the candidate presentation. All additional inference is restricted to disputed questions under the same trigger as \method{}. Unless otherwise specified, newly elicited candidate-level evidence is incorporated using the aggregation rule in Eq.~\ref{eq:represent} with $w=1$.

\subsection{Additional Computation vs.\ Re-presentation}

We first compare re-presentation against repeated inference under the original question presentation. The controls include an additional greedy generation, temperature-based sampling, and self-consistency with two additional samples. These conditions are compared with individual re-presentation views and their combination under approximately matched additional token budgets.

\begin{table*}[t]
\centering
\small
\setlength{\tabcolsep}{4.5pt}
\begin{tabular}{llrrrrr}
\toprule
\textbf{Ensemble}
& \textbf{Condition}
& \textbf{Calls}
& \textbf{Extra Tok.}
& \textbf{Acc.}
& $\boldsymbol{\Delta}$
& \textbf{Stab.\ AUROC} \\
\midrule
7--9B
& Same presentation, greedy
& 1 & 23\% & 80.74 & +0.37 & 0.614 \\
& Self-consistency, $T=0.7$
& 1 & 23\% & 80.31 & -0.05 & 0.624 \\
& Reversed
& 1 & 23\% & 81.19 & +0.81 & 0.656 \\
& Top-Two
& 1 & 20\% & 80.91 & +0.53 & 0.669 \\
& Self-consistency, two samples
& 2 & 46\% & 80.47 & +0.10 & 0.647 \\
& Top-Two + Reversed
& 2 & 43\% & \textbf{81.51} & \textbf{+1.15} & \textbf{0.697} \\
\midrule
14--32B
& Same presentation, greedy
& 1 & 15\% & 86.96 & +0.26 & 0.641 \\
& Self-consistency, $T=0.7$
& 1 & 15\% & 87.00 & +0.30 & 0.629 \\
& Reversed
& 1 & 15\% & 87.09 & +0.39 & 0.645 \\
& Top-Two
& 1 & 13\% & 87.16 & +0.46 & 0.651 \\
& Self-consistency, two samples
& 2 & 29\% & 87.13 & +0.42 & 0.646 \\
& Top-Two + Reversed
& 2 & 28\% & \textbf{87.46} & \textbf{+0.75} & \textbf{0.679} \\
\bottomrule
\end{tabular}
\caption{Controls comparing additional inference under the original presentation with alternative candidate-space views. Calls denotes additional inference calls per model on disputed questions. Extra Tok.\ reports additional combined input and generated tokens relative to the initial five-model inference. Acc.\ is overall average accuracy (\%), and $\Delta$ denotes the improvement over Stage~1. Stab.\ AUROC evaluates cross-presentation stability as a correctness indicator.}
\label{tab:compute_controls}
\end{table*}

As shown in Table~\ref{tab:compute_controls}, additional inference under the original presentation provides smaller improvements than controlled re-presentation. For the 7--9B Ensemble, an additional greedy call improves average accuracy by 0.37 percentage points, while an additional temperature-$0.7$ sample slightly decreases accuracy. In comparison, the reversed and top-two views improve accuracy by approximately 0.8 and 0.5 points, respectively. Combining both views yields the largest improvement.

A similar pattern holds for the 14--32B Ensemble. Repeated inference under the original presentation produces improvements of 0.26--0.42 points, whereas combining both re-presentation views improves average accuracy by approximately 0.76 points. These results suggest that the observed gains are not explained solely by allocating additional inference calls.

The same-presentation greedy condition should be interpreted as a control rather than an exact deterministic replay. The initial and additional generations were executed using different inference implementations, and small numerical differences may therefore affect the resulting predictions. Accordingly, this condition measures the value of another observation under the same presentation rather than exact bit-level reproducibility.

\subsection{Alternative Presentation Perturbations}

We next examine whether the benefit of re-presentation depends on a particular option-order transformation. In addition to the reversed and top-two views, we evaluate three independently sampled option permutations and a cyclic shift of the answer options. All conditions use the same selective trigger and evidence aggregation procedure.

\begin{table*}[t]
\centering
\small
\setlength{\tabcolsep}{5pt}
\begin{tabular}{llrrrr}
\toprule
\textbf{Ensemble}
& \textbf{View}
& \textbf{Flip Rate}
& \textbf{Stab.\ AUROC}
& \textbf{Acc.}
& $\boldsymbol{\Delta}$ \\
\midrule
7--9B
& Reversed     & 51.4 & 0.656 & 81.19 & +0.81 \\
& Random 1     & 47.3 & 0.649 & 81.39 & +1.01 \\
& Random 2     & 48.1 & 0.642 & 80.77 & +0.40 \\
& Random 3     & 46.9 & 0.642 & 81.41 & +1.05 \\
& Cyclic Shift & 47.7 & 0.646 & 81.14 & +0.77 \\
& Top-Two      & 43.8 & \textbf{0.669} & 80.91 & +0.55 \\
\midrule
14--32B
& Reversed     & 48.7 & 0.645 & 87.09 & +0.38 \\
& Random 1     & 47.1 & \textbf{0.653} & 87.33 & +0.63 \\
& Random 2     & 46.5 & 0.643 & \textbf{87.34} & +0.64 \\
& Random 3     & 46.0 & \textbf{0.653} & 87.27 & +0.58 \\
& Cyclic Shift & 47.7 & 0.651 & 87.20 & +0.50 \\
& Top-Two      & 42.7 & 0.651 & 87.16 & +0.46 \\
\bottomrule
\end{tabular}
\caption{Alternative presentation controls on disputed questions. Flip Rate denotes the percentage of model--question pairs whose prediction changes relative to Stage~1. Stab.\ AUROC evaluates stability as a binary correctness indicator. Acc.\ denotes overall average accuracy (\%), and $\Delta$ is the improvement over Stage~1 in percentage points.}
\label{tab:presentation_controls}
\end{table*}

Table~\ref{tab:presentation_controls} shows that several alternative presentations provide useful reliability signals and improve final accuracy. In the 7--9B Ensemble, individual random permutations yield gains ranging from 0.40 to 1.05 percentage points, while the reversed and top-two views also produce positive improvements. Similar results are observed for the 14--32B Ensemble. These findings suggest that the benefit of acquiring evidence under alternative presentations is not specific to a single fixed option permutation.

Finally, we examine the relationship between the two views used in \method{}. Their binary error indicators exhibit moderate positive correlations, with $\phi=0.34$ for the 7--9B Ensemble and $\phi=0.39$ for the 14--32B Ensemble. This incomplete overlap in prediction errors is consistent with the improvement obtained by combining the two views rather than using either view alone (Table~\ref{tab:view_ablation}).


\section{Additional Analysis of the Selective Trigger}
\label{app:trigger_analysis}

We further examine how the effectiveness of re-presentation varies with the degree of disagreement among ensemble members. Questions are grouped according to the voting pattern of the five committed single-round answers. Table~\ref{tab:trigger_analysis} reports Stage~1 accuracy, accuracy after applying re-presentation, and prediction transitions for each pattern.

\begin{table*}[!htbp]
\centering
\small
\setlength{\tabcolsep}{4pt}
\begin{tabular}{llrrrrrr}
\toprule
\textbf{Ensemble}
& \textbf{Pattern}
& $n$
& \textbf{Stage~1}
& \textbf{RePresent}
& $\boldsymbol{\Delta}$
& \textbf{W$\rightarrow$C}
& \textbf{C$\rightarrow$W} \\
\midrule
7--9B
& 5--0       & 3338 & 95.5 & 95.6 & +0.1 & 0.1 & 0.0 \\
& 4--1       & 1635 & 81.0 & 80.9 & -0.1 & 1.2 & 1.3 \\
& 3--2       & 811  & 57.1 & 63.5 & +6.4 & 13.1 & 6.7 \\
& 3--1--1    & 632  & 61.1 & 62.0 & +0.9 & 5.7 & 4.7 \\
& 2--2--1    & 434  & 45.6 & 50.5 & +4.8 & 13.1 & 8.3 \\
& 2--1--1--1 & 146  & 44.5 & 45.2 & +0.7 & 9.6 & 8.9 \\
\midrule
14--32B
& 5--0       & 4200 & 96.8 & 96.8 & +0.0 & 0.0 & 0.0 \\
& 4--1       & 1604 & 85.2 & 84.9 & -0.3 & 0.4 & 0.7 \\
& 3--2       & 508  & 57.5 & 63.2 & +5.7 & 12.8 & 7.1 \\
& 3--1--1    & 333  & 56.8 & 57.4 & +0.6 & 5.1 & 4.5 \\
& 2--2--1    & 264  & 43.6 & 47.3 & +3.8 & 13.6 & 9.8 \\
& 2--1--1--1 & 91   & 42.9 & 56.0 & +13.2 & 18.7 & 5.5 \\
\bottomrule
\end{tabular}
\caption{Accuracy (\%) and prediction transitions by single-round voting pattern. Stage~1 and RePresent denote accuracy before and after re-presentation, respectively. $\Delta$ is the accuracy difference in percentage points. W$\rightarrow$C and C$\rightarrow$W denote the percentages of questions corrected and made incorrect, respectively, within each pattern.}
\label{tab:trigger_analysis}
\end{table*}

\paragraph{Effect of disagreement.}
The gains from re-presentation are concentrated on questions with substantial disagreement. For the 7--9B Ensemble, accuracy increases by 6.4 percentage points on $3$--$2$ splits and by 4.8 points on $2$--$2$--$1$ splits. The corresponding improvements for the 14--32B Ensemble are 5.7 and 3.8 points. The $2$--$1$--$1$--$1$ configuration also shows a larger gain for the 14--32B Ensemble, although this group contains relatively few questions.

In contrast, unanimous and $4$--$1$ voting patterns exhibit little or no net benefit from re-presentation. These results suggest that strong agreement provides a useful criterion for avoiding additional inference, while questions with greater disagreement offer more opportunities for correcting Stage~1 predictions.

\paragraph{Correction and damage rates.}
Re-presentation can both correct initially incorrect predictions and change initially correct predictions to incorrect answers. For example, on $3$--$2$ splits, the 7--9B Ensemble has a correction rate of 13.1\% and a damage rate of 6.7\%, yielding a net improvement of approximately 6.4 percentage points. The corresponding rates for the 14--32B Ensemble are 12.8\% and 7.1\%. The larger correction rates explain the net gains on these disputed configurations.

\paragraph{Implications for selective inference.}
These observations support the simple agreement-based trigger used by \method{} in Eq.~\ref{eq:trigger}. By applying additional inference only when the largest agreement group contains at most $K-2$ models, the framework targets questions where re-presentation is more likely to improve the aggregate prediction. This analysis supports the chosen heuristic but does not establish its optimality relative to alternative confidence-based or learned triggering policies.

\section{Implementation Details}

\subsection{Inference Prompt}
\label{app:prompt}

All ensemble members use the same prompt template for
multiple-choice reasoning. The template requests a short
chain-of-thought response followed by a final answer letter.
The question and answer options are substituted for the
corresponding placeholders. For re-presentation, the same
template is used with the modified candidate set and
option ordering.

\begin{quote}
\small
\ttfamily
Answer the following multiple-choice question.\\
Give a short chain-of-thought in a few steps, then output the Final Answer.\\[4pt]
Question: \{question\}\\
Options:\\
A. \{option A\}\\
B. \{option B\}\\
...\\[4pt]
Use exactly this format:\\
Chain-of-Thought:\\
1) ...\\
2) ...\\[4pt]
Final Answer: <one letter>
\end{quote}

\subsection{Computational Resources and Configuration.}
All experiments were implemented using PyTorch and Hugging Face Transformers. All experiments were conducted using NVIDIA H100 NVL GPUs.The evaluated models range from 7B to 32B parameters. Inference costs are reported in terms of input, output, and total token usage, normalized by the initial five-model inference.

\end{document}